\documentclass[aps,10pt,pra,twocolumn,superscriptaddress,showpacs,amsmath,amssymb,nobibnotes,floatfix]{revtex4-1}

\usepackage{graphicx}
\usepackage{mathptmx}         
\usepackage{dcolumn}          
\usepackage[mathlines]{lineno}
\usepackage{bm}               
\usepackage{amsmath}
\usepackage{booktabs}
\usepackage{multirow}
\begin{document}

\title{Maximum likelihood \textit{q}-estimator reveals nonextensivity regulated by extracellular potassium in the mammalian neuromuscular junction}

\author{A. J. da Silva}
\email[]{adjesbr@gmail.com}
\affiliation{Departamento de F\'isica, Instituto de Ci\^encias Exatas, Universidade Federal de Minas Gerais,
CEP 31270-901, Belo Horizonte, Minas Gerais, Brazil}

\author{M. A. S. Trindade}
\affiliation{Departamento de Ci\^encias Exatas e da Terra, Universidade do Estado da Bahia, CEP 48040-210, Alagoinhas, Bahia, Brazil}

\author{D. O. C. Santos}
\affiliation{Departamento de Bioqu\'imica e Imunologia, Instituto de Ci\^encias Biol\'ogicas, Universidade Federal de Minas Gerais,CEP 31270-910, Belo Horizonte, Minas Gerais, Brazil}

\author{R. F. Lima}
\affiliation{Departamento de Fisiologia e Farmacologia, Faculdade de Medicina, Universidade Federal do Cear\'a, CEP 60430-270, Fortaleza, Cear\'a, Brazil}

\date{\today}

\begin{abstract}
Recently, we demonstrated the existence of nonextensivity in neuromuscular transmission [Phys. Rev. E 84, 041925 (2011)]. In the present letter, we first obtain an estimator, the maximum likelihood \textit{q}-estimator, applied to calculate the scale factor ($\alpha$) and the \textit{q}-index of \textit{q}-gaussian distributions. Next, we used our theoretical findings to analyze spontaneous miniature end plate potentials in electrophysiological recordings from neuromuscular junction. These calculations were performed in both normal and high extracellular potassium concentration, $[K^{+}]_{o}$. This protocol was used to test the validity of the \textit{q}-index in electrophysiological conditions closely resembling physiological stimuli. The analysis showed a significant difference between \textit{q}-indexes in high and normal $[K^{+}]_{o}$, where the magnitude of nonextensivity was increased. Our letter provides a general way to obtain the best \textit{q}-index from the \textit{q}-Gaussian distribution function. It also expands the validity of Tsallis statistics in a realistic physiological stimulus condition. We also discuss in detail physical and physiological implications of these findings.
\end{abstract}


\maketitle

\section{Introduction}
The neuromuscular junction (NMJ) is the structure responsible for communicating electrical impulses from the motor neuron to the skeletal muscle to signal contraction \cite{hughes}. The specialized terminal formed by the NMJ constitute a prototypical example of a chemical synapse. Using this structure, physiological and pharmacological investigations have explored many characteristics of how chemical communication is accomplished. Perhaps, the most notable finding in this preparation was announced by Fatt and Katz in the early 1950s, using the frog NMJ \cite{katz1952}. They detected an intermittent form of biological noise, the spontaneous miniature end plate potential (MEPP), later attributed to the spontaneous release of acetylcholine from the nerve ending. As a consequence, in 1954, the first quantitative description about this phenomenon was developed, inaugurating a vast field in Neurophysiology \cite{katz1954}. The 1954 study associated the statistics of spontaneous release to Gaussian and Poisson models (binomial statistics). However, subsequent reports for different neuromuscular preparations suggest that spontaneous transmitter release seems to violate a Gaussian process also departing from the Poisson prediction \cite{wernig,washio,kloot99,lowen,takeda}. The data suggest that spontaneous quantal release is not a product of independent events, but rather it depends on its previous history or it shows long-range correlations.

Recently, we demonstrated the existence of long-range correlations associated to spontaneous release at the NMJ of mouse diaphragm \cite{adriano}. We showed that the miniature end plate potential (MEPP) can be described by \textit{q}-Gaussian statistics, providing an alternative approach to quantal analysis. The formalism of Nonextensive Statistics proposed by Tsallis is promising in describing the nature of neurotransmitter discharge, as offers a concise explanation for the interactions in the ending terminal machinery, responsible for long-range correlations \cite{harlow}. Nonextensive statistics describes systems in which the entropy is not proportional to system size, a property commonly present in complex systems displaying long-range interactions (or correlations) or out of equilibrium. Nowadays, \textit{q}-distributions have been widely employed in many complex systems in Economics, Physics and Chemistry, due to its peculiar ability to show heavy-tails and model power law phenomena \cite{tsallis2}. Despite its widespread verification, nonextensive statistics remains scarcely applied to Neurobiology \cite{zhang,capurro}.

In Statistics, the method of Maximum Likelihood Estimation (MLE) has been used to obtain statistical properties of distributions \cite{bustamante}. MLE has also been widely used to address physical problems such as drift and diffusion in time series, multifractal random walks, and recognition of $\alpha$-stable Levy distributions \cite{rypdal,kleinhans,gonchar}. This method has also been successfully employed, in conjunction with nonextensive statistics, to analyze self-organized criticality in Ehrenfest dog-flea model \cite{bakar}. With respect to Neurophysiology applications, MLE has been adopted to address fine aspects of quantal release in several NMJs \cite{cooper}. Recently, MLE has been successfully applied to nonextensive statistics allowing more accuracy in the extraction of statistical parameters \cite{shalizi}.  Based on \textit{q}-calculus, a \textit{q}-version of the Error Law was developed using MLE \cite{suyari}. Recently, estimators satisfying a generalized version of Cramer-Rao inequality, such as a \textit{q}-version of Fisher information, were presented
\cite{bercher}. Further, ML\textit{q}E estimators were obtained and their properties investigated with asymptotic analysis and also with computational simulations \cite{bercher,ferrari}. MLE has been useful to uncover refined aspects underlying the non-uniform release of neurotransmitters and interactions among release sites \cite{smith}. This method was used to understand statistical behavior in short term plasticity at the lobster NMJ \cite{worden}. Therefore, with such advances, \textit{q}-statistics, combined with MLE, emerges as a relevant formalism to examine real-world data.

Inspired by Tsallis's ideas, in this report we introduce calculations to obtain generalized estimators for the scale factor $\alpha$ and the \textit{q}-index from NMJ recordings using ML\textit{q}E estimators. The scale factor regulates the width (variance) of \textit{q}-Gaussian distributions and the \textit{q}-index or entropic index is a measure of nonextensivity. We also investigate the influence of nerve hyperexcitability, stimulated by high $[K^{+}]_{o}$  in the magnitude of nonextensivity at the NMJ. In the literature, the role of calcium in binomial statistics has been considered in different neuromuscular junctions, but the role of potassium remains obscure even in classical statistical descriptions. We believe that the electrophysiological patterns generated by high $[K^{+}]_{o}$ constitute a system in which it is possible to explore the role of this ion in modulating long-range correlations. Beyond that, ML\textit{q}E estimation can emerge as a method to identify the presence of nonextensivity in pathological cases related to elevation of $[K^{+}]_{o}$. Additionally, we propose a closer relationship between physiological mechanisms and statistical models associated to neuroplasticity. For that, we use nonextensive statistics to establish a bridge between long-range correlations and synaptic transmission.

\section{Materials and Methods}

\subsection{Theoretical methods}
Thus, inspired by general properties of \textit{q}-algebra, we calculate, analytically, \textit{q}-estimators for $\alpha$ while the \textit{q}-index is numerically calculated. The strategy in our calculation is to assume \textit{q}-estimators of \textit{q}-products of \textit{q}-functions. From this assumption we expect to extract a generalized expression make up by \textit{N} estimators for \textit{N} random variables that obey \textit{q}-Gaussians distributions with distinct $\alpha$. Our main motivation for obtaining estimators lies in the fact that researchers have routinely fitted \textit{q}-Gaussian distributions to their experimental data, when the use of estimators would increase the accuracy of results for a measurement of nonextensivity degrees. To guarantee that the probability is at maximum we perform a second derivative test for $\alpha$.  With this probability, the \textit{q}-index is numerically estimated. Thus, this \textit{q}-estimator could be used to evaluate the $\alpha$ and the \textit{q}-index from any biological data set, where the motivation is uncover long range correlation.

\subsection{Experimental methods}
In our experimental design we elevated $[K^{+}]_{o}$ for the following reasons. First, this ionic manipulation is credited to closely mimic physiological stimulation in many neuronal systems \cite{grohovaz}. Second, the effect of the external and internal potassium concentration over the membrane potential in muscle preparations is well known \cite{adrian}. At NMJ, high $[K^{+}]_{o}$ initiates strong membrane depolarization accompanied by acceleration of the MEPP rate \cite{glavinovic}. Third, a number of studies have correlated morphological cellular modifications evoked by the accumulation of $[K^{+}]_{o}$       \cite{sykova}. Finally, we investigate a possible regulation of nonextensivity by the modulatory effects of ions on MEPP rate and thus on statistical parameters.

The hemidiaphragm is a muscle that separates the thoracic from the abdominal cavity and presents several advantages. For example, the use of the NMJ is readily justified by its easy identification, simplicity to dissect and to extract the muscles, and by its stereotyped spontaneous electrophysiological activity. All experimental procedures in the present work were approved by the Animal Research Committee (CETEA - UFMG, protocol 073/03) \cite{caboco}. Wild-type adult mice were euthanized by cervical dislocation. Diaphragms were quickly removed and inserted in a physiological control solution containing (in mM): NaCl (137), NaHCO$_{3}$ (26), KCl (5), NaH$_{2}$PO$_{4}$ (1.2), glucose (10), CaCl$_{2}$ (2.4), and MgCl$_{2}$ (1.3). pH was adjusted to 7.4 after gassing with 95\% O$_{2}$ and 5\% CO$_{2}$. In the experiments with high $[K^{+}]_{o}$ (25 mM), in bath solution, sodium concentration was adjusted to maintain the osmotic equilibrium. Muscles were maintained in solution at least 30 minutes before the beginning of the electrophysiological recordings allowing a recover from the mechanic trauma of their extraction. Tissues were transferred to a recording chamber continuously irrigated with fresh fluid at a rate of 2-3 ml/min at room temperature ($T = 24 \pm 1^\circ{}C$). Standard intracellular recording technique was used to monitor the frequency of spontaneous MEPP by inserting a micropipette at the chosen muscle fiber. Borosilicate glass microelectrodes had resistances of 8-15 M$\Omega$ when filled with 3 M KCl. A single pipette was inserted into the fiber near the end-plate region as guided by the presence of MEPP with rise times $<$ 1 ms. Amplitudes and areas were computed at both control and 25 mM of $[K^{+}]_{o}$ in the bath solution, where at least 900 MEPP were used to evaluate the \textit{q}-index. This mandatory precondition supplied a sufficient number of events required for a rigorous theoretical analysis. We harvested 16 recordings in different fibers from 8 animals.

\section{Results}

\subsection{\textit{q}-likelihood estimator function from a \textit{q}-Gaussian distribution}

In this section we present the calculation of maximum likelihood estimators for the scale factor, $\alpha$ and the \textit{q}-index. We start by defining the \textit{q}-Gaussian function as:
   \begin{equation}
   f_{q}(x)=\frac{\sqrt{\alpha}}{C_{q}}\ \exp_{q}(-\alpha x^{2}),
   \end{equation}
where,
   \begin{equation}
   \exp_{q}(x)=[1+(1-q)x]^{1/(1-q)},
   \end{equation}
and the normalization constant:
\begin{equation}
 C_q=\left\{	\begin{array}{ll}
   \dfrac{2\sqrt{\pi} \ \Gamma\left( \dfrac{1}{1-q}\right)}{(3-q)\sqrt{1-q} \ \Gamma\left(\dfrac{3-q}{2(1-q)}\right)},&  -\infty<q<1 \\
		\vspace{0.1cm}
    \sqrt{\pi},&  \ \ q=1 \\
		\vspace{0.1cm}
    \dfrac{\sqrt{\pi} \ \Gamma\left(\dfrac{3-q}{2(q-1)}\right)}{\sqrt{q-1} \ \Gamma\left( \dfrac{1}{1-q}\right)},&   1<q<3 \\
  \end{array}	\right.
\end{equation}
		
${C_{q}}$ domain was assumed from results of our previous report \cite{adriano}, where the \textit{q}-index belongs to the region $1<\textit{q}<3$. To perform a rigorous estimate of the parameters $\alpha$ and \textit{q} from a \textit{q}-Gaussian distribution, consider $x_{k_{1}}, ..., x_{k_{n}}$ a random sample of size $n_{k}$ of the random variable $X_{k}$ with a $q$-density function$f_{q}(x_{k},\theta_{k})$ and $\theta$ $(((\theta$=$\theta_{k_{1}},...,q,...,\theta_{k_{n}}))\in\Theta$
$\Theta$ is defined as the parameter space. The \textit{q}-likelihood as a function of $\theta$ is:
\begin{equation}
     l_{q}(\theta_{k}, x_{k})=\bigotimes_{j=1\phantom{i}q}^{n_{k}} f_{q}(x_{k_{j}}|\theta_{k_{j}})
\end{equation}

The \textit{q}-product, $\bigotimes_{q}$ is written as:
     \begin{equation}
     x {\otimes}_{q} y\equiv[x^{1-q}+y^{1-q}-1]_{+}^{\frac{1}{1-q}}
     \end{equation}
With  $(x,y) > 0$ and $[A]_{+}\equiv max[A,0]$. For $N$ random variables $X_{1},...,X_{N}$ and their parameters $\theta_{1}, ...,\theta_{N}$ we have the $q$-likelihood function:
   \begin{eqnarray}
    L_{q}(\theta_{1},...,\theta_{N}, x_{1},...,x_{N})
		&=&\bigotimes_{i=1\phantom{i}q}^{N_{q}} l_{q}(x_{i}|\theta_{i}) \nonumber \\
		&=&\bigotimes_{i=1\phantom{i}q}^{N_{q}}\left(\bigotimes_{j=1\phantom{i}q}^{n_{i}} f_{q}(x_{i_{j}}|\theta_{i_{j}})\right)
   \end{eqnarray}

The maximum \textit{q}-likelihood estimator is:
     \begin{equation}
     \tilde{\theta_{i}^{q}} = \arg \max_{\theta_{i}^{q}\in\Theta}\ln_{q}(L_{q}(\theta_{1},...,\theta_{N}, x_{1},...,x_{N})),
		\end{equation}
where $\ln_{q}(x)=(x^{1-q}-1)/(1-q)$.

For \textit{q}-Gaussian functions follows the \textit{ q}-likelihood estimator, $L_{q}= L_{q}(\alpha_{1},...,\alpha_{N}, x_{1}, ...,x_{N})$:
\begin{equation}
   \begin{split}
L_{q}=&\left[\left(\frac{\sqrt{\alpha_{1}}}{C_{q}}\right)\exp_{q}(-\alpha_{1} x_{1_{1}}^2)\right]...\\
		 &\otimes_{q}\left[\left(\frac{\sqrt{\alpha_{1}}}{C_{q}}\right)\exp_{q}(-\alpha_{1} x_{1_{n_{1}}}^2)\right]...\\
     &\otimes_{q}\left[\left(\frac{\sqrt{\alpha_{N}}}{C_{q}}\right)\exp_{q}(-\alpha_{N} x_{N_{1}}^2)\right]...\\
		 &\otimes_{q}\left[\left(\frac{\sqrt{\alpha_{N}}}{C_{q}}\right)\exp_{q}(-\alpha_{N} x_{N_{n_{N}}}^2)\right]
		\end{split}
  \end{equation}
		
Using the identity: \begin{equation}\exp_{q}(x)\exp_{q}(y)=\exp_{q}[x\pm y\pm(1-q)xy]\end{equation} it can be shown that:
   \begin{equation}	\label{eq:loglike}
	  \begin{split}
     L_{q}=&\exp_{q} \biggl\{ \sum_{j=1}^{N} \biggl\{ (-\alpha_{j})\left[1+(1-q)\ln_{q}\left(\frac{\sqrt{\alpha_{j}}}{C_{q}}\right)\right]   \biggr.\biggr.\\
	   & \times\biggl. \biggl.\sum_{i_{j}=1}^{n_{j}}x_{j_{i_{j}}}^{2}+n_{j}\ln_{q}\left(\frac{\sqrt{\alpha_{j}}}{C_{q}}\right)		\biggl.\biggr\} \biggl.\biggr\}
	\end{split}
   \end{equation}
	
\subsection{\textit{q}-likelihood estimate for the scale factor $\alpha$}

To determine the maximum \textit{q}-likelihood estimate of the scale factor $\alpha$ we calculate $\ln_{q}$ of the previous expression(\ref{eq:loglike}). Then, the derivative as a function of $\alpha_{1},  ..., \alpha_{N}$ results in the ollowing equation:
\begin{equation}
\begin{split}
   D_{\alpha_{j}}^{'}[\ln_{q}(L_{q})]=&-\left(\frac{3-q}{2C_{q}^{1-q}}\right)\ \alpha_{j}^{\frac{1-q}{2}}\ \sum_{i=1}^{n_{j}}x_{i}^2  \\
	&+\frac{n\alpha_{j}^{\frac{-1-q}{2}}}{2C_{q}^{1-q}}, \ j = 1, ...,N
	\end{split}
  \end{equation}
which has as a stationary solution:
   \begin{equation}
   \alpha_{j} = \frac{n_{j}}{(3-q) \displaystyle\sum_{i_{j}=1}^{n_{j}}x_{j_{i}}^2}, \ j = 1, ...,N
    \label{est}
   \end{equation}
and second derivative:
   \begin{equation}
	\begin{split}
   D_{\alpha_{j}}^{''}[\ln_{q}(L_{q})]=&\frac{-(3-q)(1-q)}{4C_{q}^{-1-q}} \alpha_{j}^{\frac{1-q}{2}}\sum_{i=1}^{n_{j}}x_{i}^2  \\
	&- \frac{(1+q)}{4C_{q}^{1-q}\alpha_{j}}n_{j}\alpha_{j}^{\frac{-1-q}{2}}, \ j = 1, ..., N
	\end{split}
   \end{equation}

 It is easily demonstrated that $\alpha_{1}, ..., \alpha_{N}$, given by the starionary solution(\ref{est}) are maxima in the region $1<\textit{q}<3$, as
   \begin{equation}
   (q-1)\frac{n_{1}}{\alpha_{j}} < (q+1)\frac{n_{j}}{\alpha_{j}},  j = 1, ..., N
   \end{equation}
To our purposes we need only one  \textit{q}-Gaussian distributed random variable. In that case:
     \begin{eqnarray} \label{eq:loglike1}
     L_{q}(\theta,x)&=&\left[\left(\frac{\sqrt{\alpha}}{C_{q}}\right)\exp_{q}(-\alpha x_{1}^2)\right]  \nonumber\\
		&\otimes_{q}&...\nonumber\\
		&\otimes_{q}&\left[\left(\frac{\sqrt{\alpha}}{C_{q}}\right)\exp_{q}(-\alpha x_{n}^2)\right]
     \end{eqnarray}	
As a result:
  \begin{equation}
	\begin{split}
   L_{q}(\theta,x)=&
	\exp_{q}\biggl\{(-\alpha) \left[1+(1-q)\ln_{q}\left(\frac{\sqrt{\alpha}}{C_{q}}\right)\right]\sum_{i=1}^{n}x_{i}^2 \biggr.  \\
	&
	\biggr. +n\ln_{q}\left(\frac{\sqrt{\alpha}}{C_{q}}\right) \biggr\}
	\end{split}
   \label{LL}
   \end{equation}
		
Calculating $\ln_{q}$ of the expression(\ref{LL}) and deriving it in relation to $\alpha$ we get:
   \begin{equation}
   D_{\alpha}^{'}[\ln_{q}(L_{q}(\theta,x)]=-\left(\frac{3-q}{2C_{q}^{1-q}}\right)\alpha^{\frac{1-q}{2}}
	\sum_{i=1}^{n}x_{i}^2+\frac{n\alpha^{\frac{-1-q}{2}}}{2C_{q}^{1-q}}
   \end{equation}
The stationary solution is a particular case of equation(\ref{est}) given by:
   \begin{equation}
   \alpha = \frac{n}{(3-q)\displaystyle\sum_{i=1}^{n}x_{i}^2}
   \end{equation}

   In the limit $q\rightarrow 1$, we recover the usual Gaussian case. This particular result is similar to the obtained by Hasegawa and Arita \cite{hasegawa}. The second derivative in relation to $\alpha$ is given by:
   \begin{equation}\begin{split}
   D_{\alpha}^{''}[\ln_{q}(L_{q}(\theta,x)]=&\dfrac{-(3-q)(1-q)}{4C_{q}^{-1-q}} \alpha^{\frac{1-q}{2}} \\
	&\times\displaystyle\sum_{i=1}^{n}x_{i}^2 - n\dfrac{(1+q)}{4C_{q}^{1-q}\alpha}\alpha^{\frac{-1-q}{2}}
   \end{split}\end{equation}
	
It is also possible to generalize the estimator for $N$ random variables with the same $\alpha$. In this case:
\begin{equation}
	\begin{split}
   L_{q}=&\exp_{q}(-\alpha)\left[1+(1-q)\ln_{q}\left(\frac{\sqrt{\alpha}}{C_{q}}\right)\right] \\
  &\times\sum_{i_{j}=1}^{n_{j}}x_{1_{i_{j}}}^{2}+n_{j}\ln_{q} \left(\frac{\sqrt{\alpha}}{C_{q}}\right)
	\end{split}
   \end{equation}
	Taking $\ln_{q}$ of the previous expression and deriving it in relation to $\alpha$ results:
	
\begin{equation}
\begin{split}
   D_{\alpha}^{'}[\ln_{q}(L_{q})]=&\sum_{j=1}^{N} \biggl\{ -\left(\frac{3-q}{2C_{q}^{1-q}}\right)\ \alpha^{\frac{1-q}{2}}\sum_{i=1}^{n_{j}}x_{i}^{2} \biggr. \\
	&+  \biggr. n_{j} \frac{\alpha^{\frac{-1-q}{2}}}{2C_{q}^{1-q}}  \biggr\}
\end{split}
\end{equation}
Finally, the stationary solution is:
   \begin{equation}
   \alpha=\frac{\left(n_{1}+...+n_{N}\right)}
	{(3-q)\left(\displaystyle \sum_{i_{1}=1}^{n_{1}} x_{1_{i}}^{2} +...+\displaystyle\sum_{i_{N}=1}^{n_{N}} x_{1_{i}}^2 \right)}
   \end{equation}
Following a procedure similar to the previous one we can show that the above solution is a maximum.

\subsection{\textit{q}-likelihood estimate for the \textit{q}-index}

The maximum \textit{q}-likelihood estimate of the \textit{q}-index is obtained by calculating the $\ln_{q}$ of the equation(\ref{eq:loglike1}) and deriving it as a function of \textit{q}. As a result:
\begin{equation}\label{deriv1}
\begin{split}
    D_{q}^{'}[\ln_{q}(L_{q})]=&-\frac{n}{{(1-q)}^{2}}+G_{1}(q)\left\{\frac{1}{{(1-q)}^{2}} \right. \\
  &\left. - \frac{2G_{2}(q) + 1}{(3-q)(1-q)}+\frac{2G_{3}(q)}{(3-q)}  \right\}
	\end{split}
 \end{equation}	
with $G_{1}(q)=n\alpha^{\frac{1-q}{2}}/C_{q}^{1-q}$, $G_{2}(q)=(\ln \alpha/2)- \ln C_{q}$ and $G_{3}(q)=\Psi\left(y-\frac{1}{2}\right)-\Psi(y)$, where $y=1/(q-1)$ and $\Psi(y)$ is the digamma function. To find out if the \textit{q}-index is a maximum we also calculate the second derivative:

\begin{equation}\label{deriv2}
\begin{split}
D_{q}^{''}[\ln_{q}(L_{q})]=&- \frac{2 n}{{(1-q)}^{3}}+
G_{1}(q)\biggl\{-\frac{1}{2}+\frac{(1-q)}{2(3-q)} \biggr.     \\
&\biggl.-G_{2}(q)-(1-q)G_{3}(q)\biggr\}\biggl\{\frac{1}{{(1-q)}^{2}}  \biggr.  \\
&-\frac{2}{(3-q)(1-q)}\left[G_{2}(q)+\frac{1}{2}\right]      \\
&\biggl.+\frac{2}{(3-q)}\left[\frac{1}{(2-q)} + G_{4}(q)\right]\biggr\}
	\end{split}
  \end{equation}	
where $G_{4}(q)=\Psi^{'}\left(y-\frac{1}{2}\right)-\Psi^{'}(y)$.

\begin{figure}[ht]
\includegraphics[width=8cm]{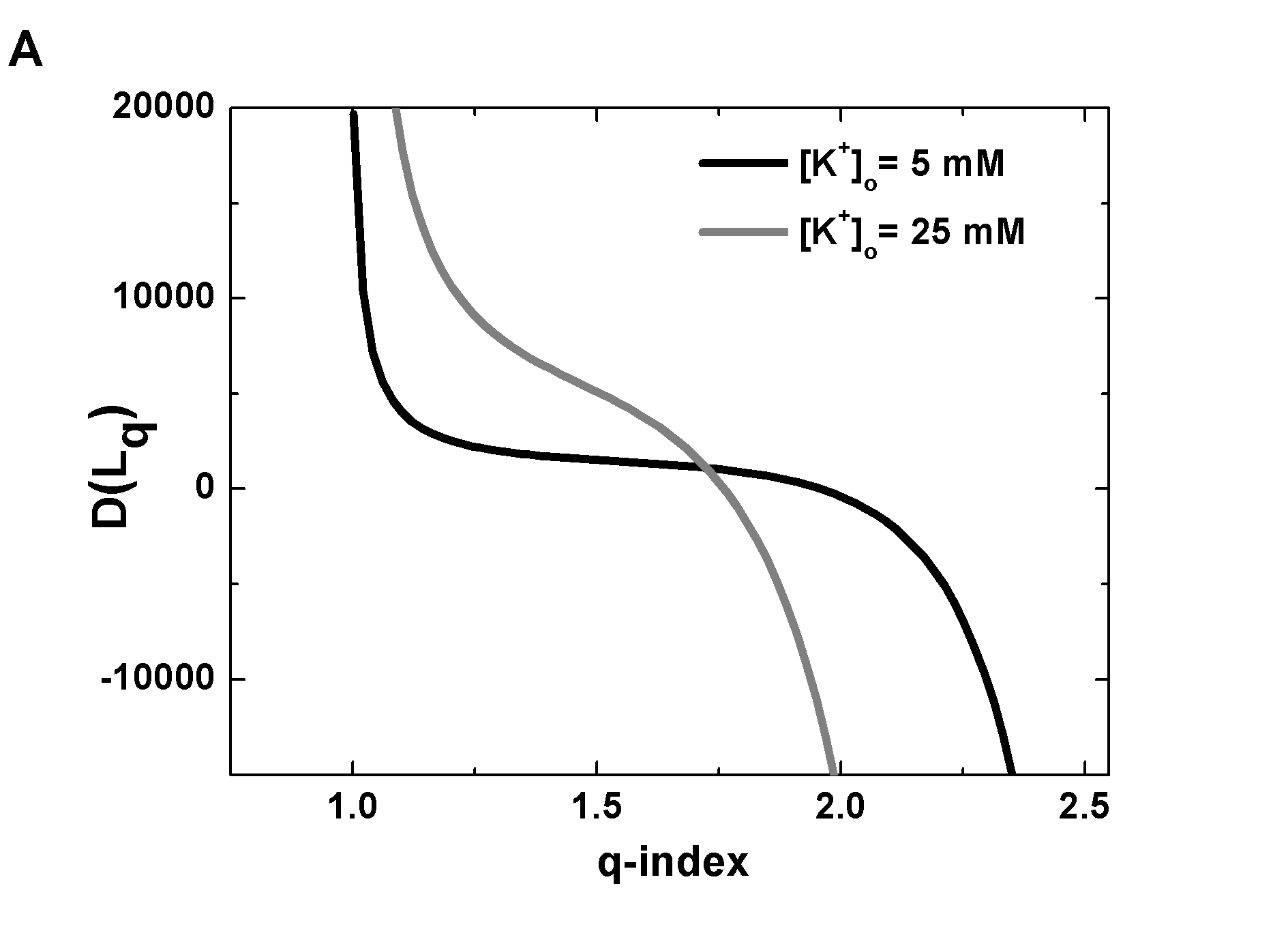}
\includegraphics[width=8cm]{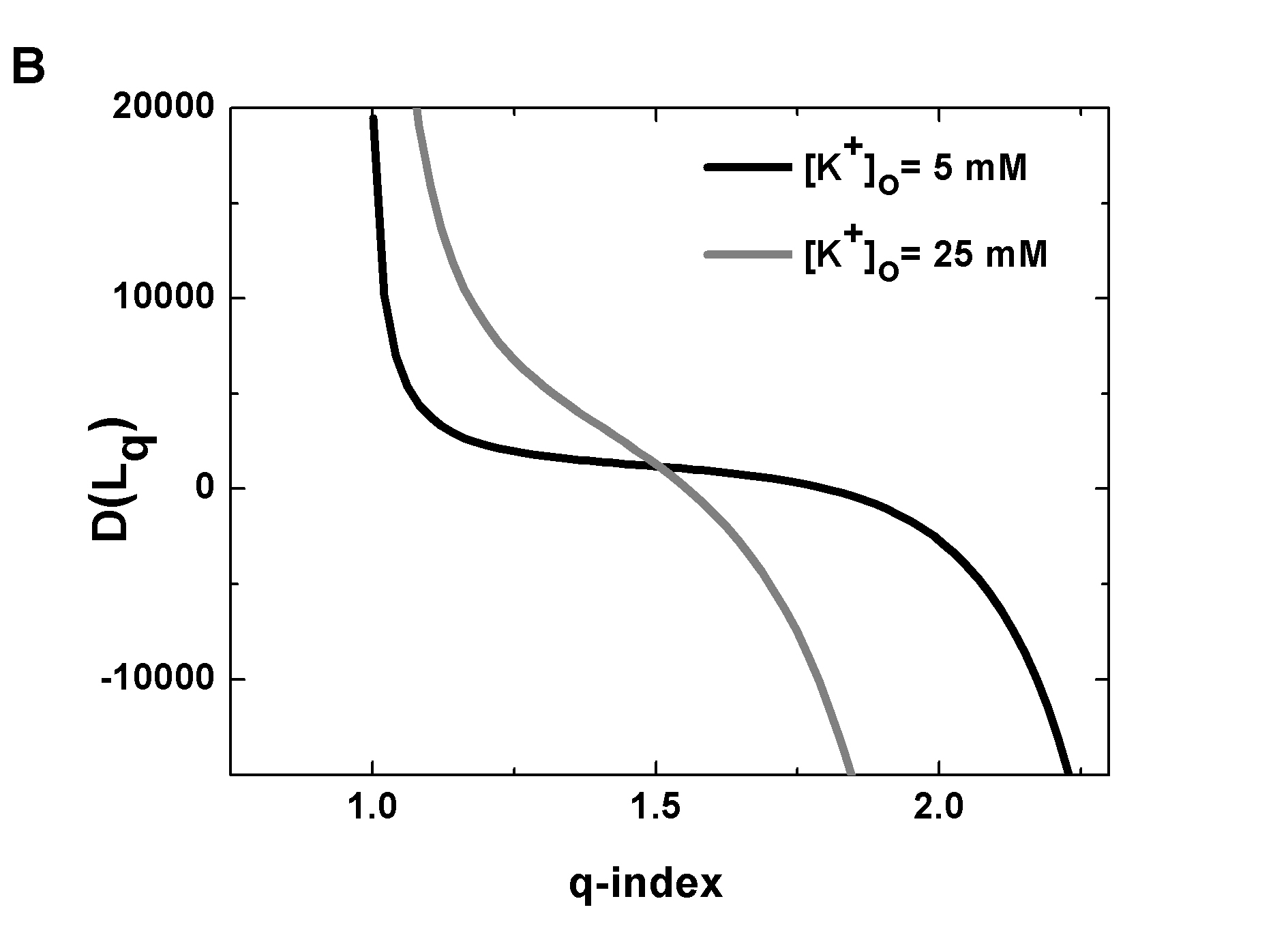}
\caption{\label{figure1} First derivative of the likelihood function, eq.(~\ref{eq:loglike}) as a function of $q$ for MEPP amplitude (a) and area (b).}
\end{figure}

Equation(\ref{deriv1}) and its second derivative (equation(\ref{deriv2})) were numerically solved to obtain the maximum likelihood estimate of the \textit{q}-index, based on the likelihood function, eq.(\ref{eq:loglike}). Figure~\ref{figure1} shows the behavior of the first derivative, eq.(\ref{deriv1}), as a function of the \textit{q}-index for amplitudes and areas of MEPPs, respectively. Note that the zero-crossings of the first derivative occur at distinct \textit{q}-index values for different $[K^{+}]_{o}$ in both figures (see also table~\ref{table2}).

\subsection{Experimental results: high $[K^{+}]_{o}$ changes nonextensivity in NMJ}
\begin{figure}[ht]
\includegraphics[width=8cm]{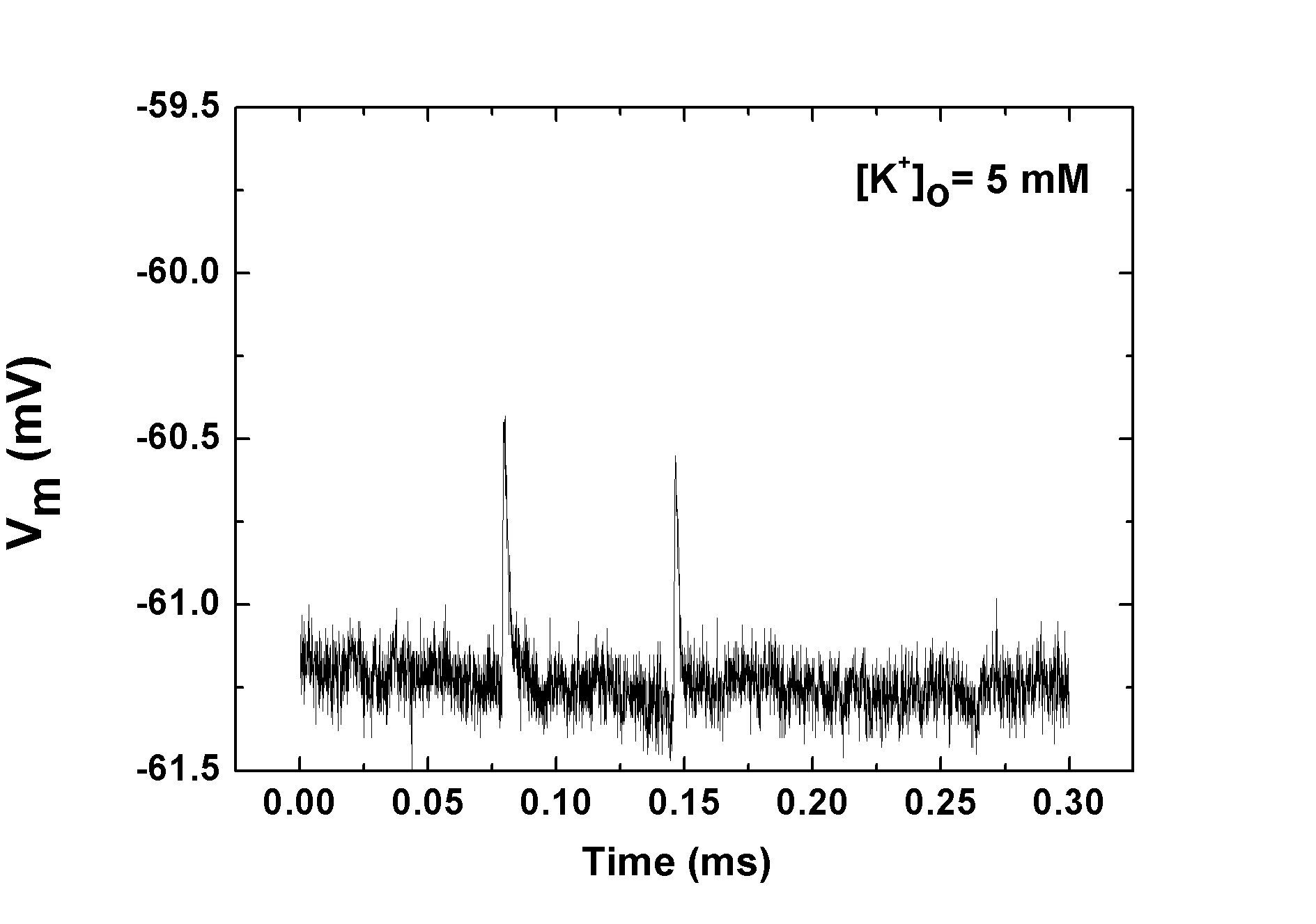}
\includegraphics[width=8cm]{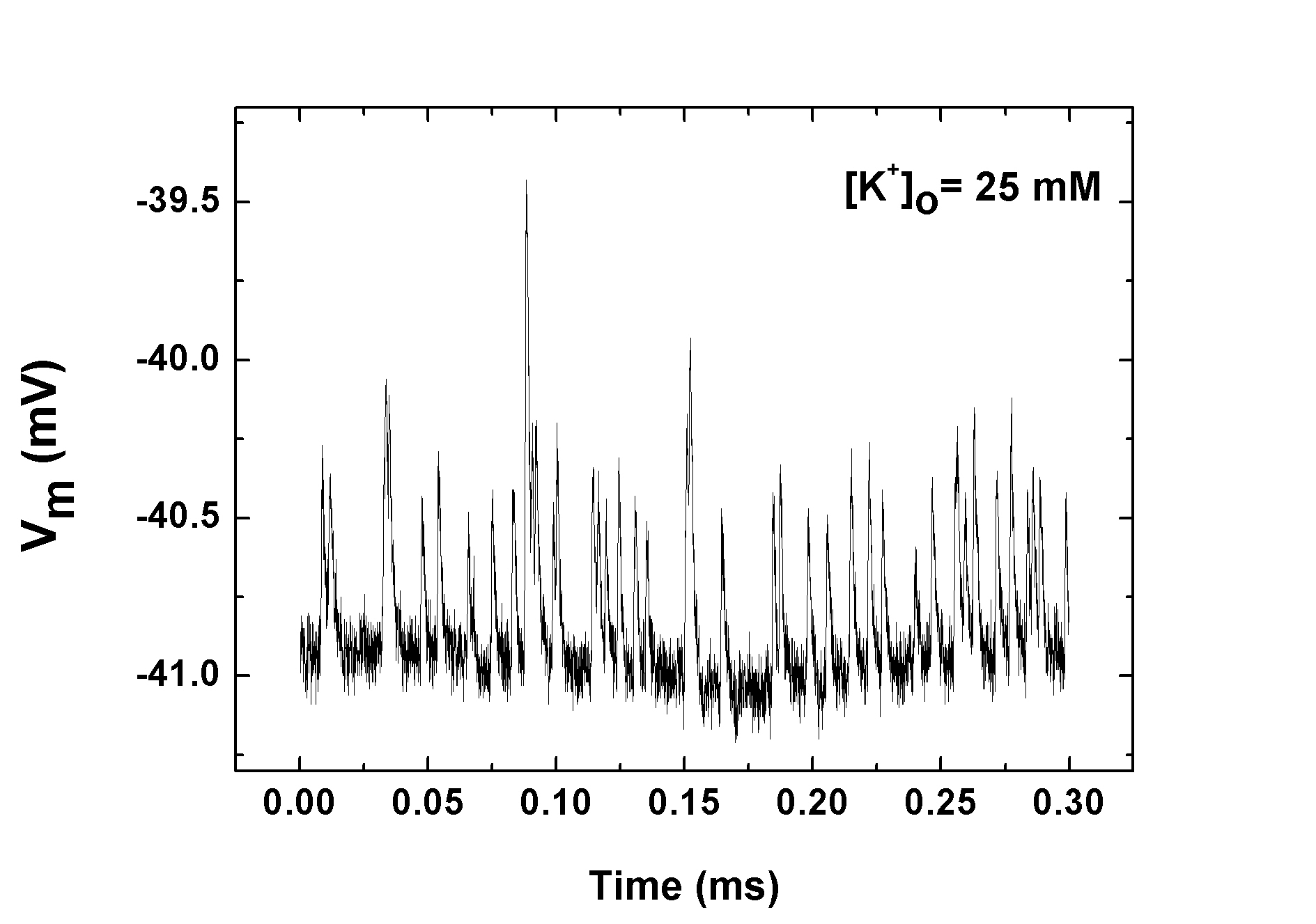}
\caption{\label{figure2} Representative recording segments of MEPP. Top: time series interval during control solution ($[K^{+}]_{o}$= 5 mM). Bottom: typical segment with elevated $[K^{+}]_{o}$ (25 mM).}
\end{figure}

The changes in MEPP rate, amplitude and area, when the NMJ was submitted to  $[K^{+}]_{o}$, were analysed and the effects on the \textit{q}-index were evaluated. We chose to include MEPP areas in conjunction with the usual amplitude analysis because areas are less susceptible to errors related to micropipette
positioning and nerve ending geometry \cite{klootreview}. The transition from low to high discharge rate evoked by high $[K^{+}]_{o}$ permits to investigate the behavior of the nonextensive index. However, prior to the main analysis itself, membrane potencial and MEPP rate were computed. In addition, we also recorded membrane potential depolarizations accompanied by an standing increase in MEPP rate, in agreement with other studies \cite{takeuchi,parsons}. These results are in concordance with the expected electrophysiological pattern response due the enhancement of  $[K^{+}]_{o}$. Thus, from these control analysis, we were confident to perform nonextensive analysis. These experimental data are sumarized in table~\ref{table1}.
\begin{table}[ht]
\caption{\label{table1} Summary of electrophysiological profile for both $[K^{+}]_{o}$.}
\begin{tabular}{@{}ccc@{}}
\hline
\multicolumn{1}{l}{\multirow{2}{*}{}} & \multicolumn{2}{c}{Concentration}           \\ \cmidrule(l){2-3}
\multicolumn{1}{l}{}                  & 5 mM                & 25 mM                  \\ \midrule
Frequency (Hz)                        & $0.81 \pm 0.48$     & $141.0 \pm 52.8$       \\
Number of MEPP                        & $682.4 \pm 323.7$ & $60324.9 \pm 28314.4$ \\
Membrane Potential (mV)               & $-70.75 \pm 8.45$   & $-39.62 \pm 4.60 $
\end{tabular}
\end{table}
High $[K^{+}]_{o}$ induces an overall increase in excitability, inducing depolarization of the resting potential, along with an expressive increase in MEPP frequency as shown in figure~\ref{figure2}. In this figure, representative recording segments of MEPP are shown, taken from normal and high $[K^{+}]_{o}$ in the bath.

\begin{figure}[ht]
\includegraphics[width=8cm]{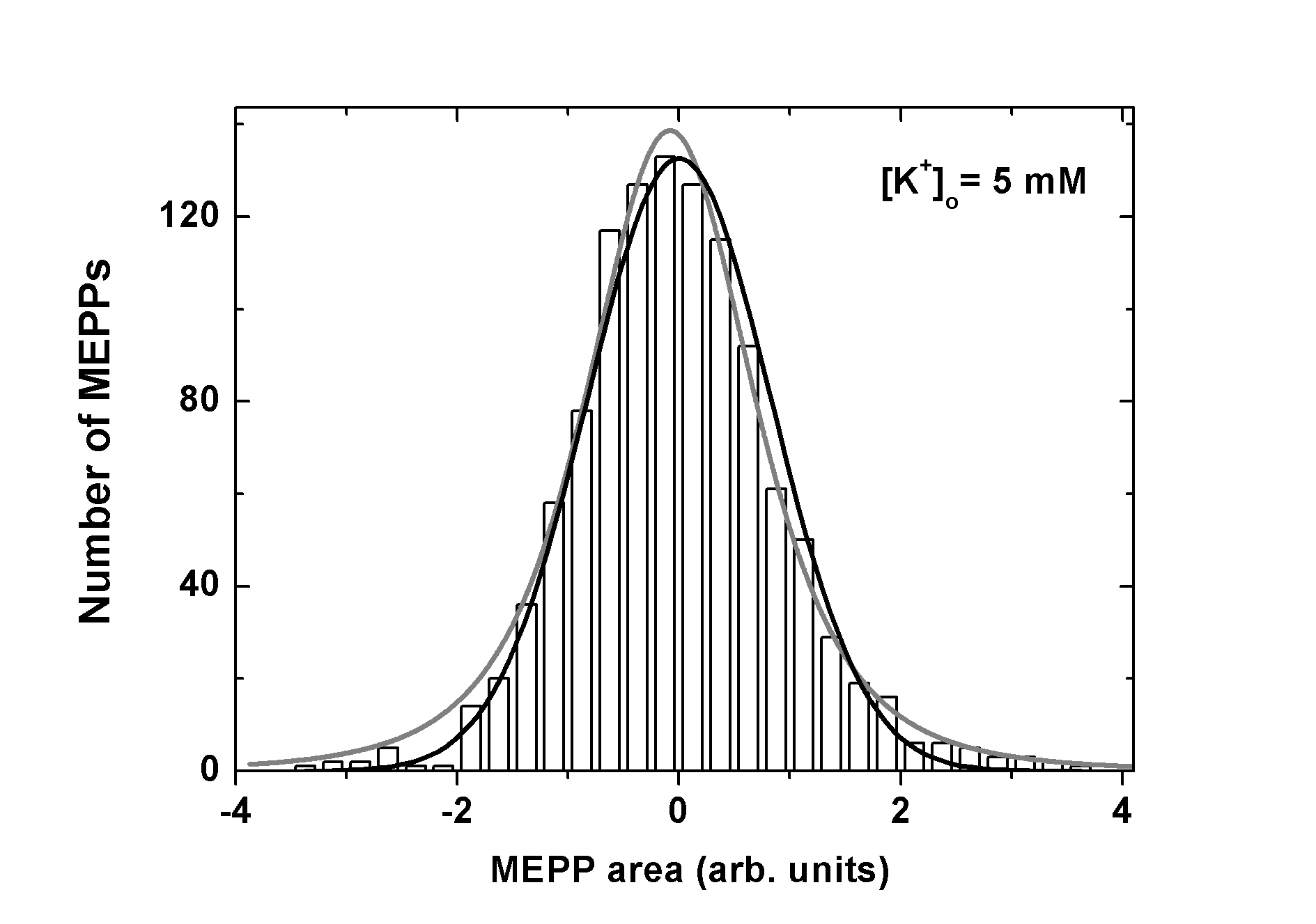}
\includegraphics[width=8cm]{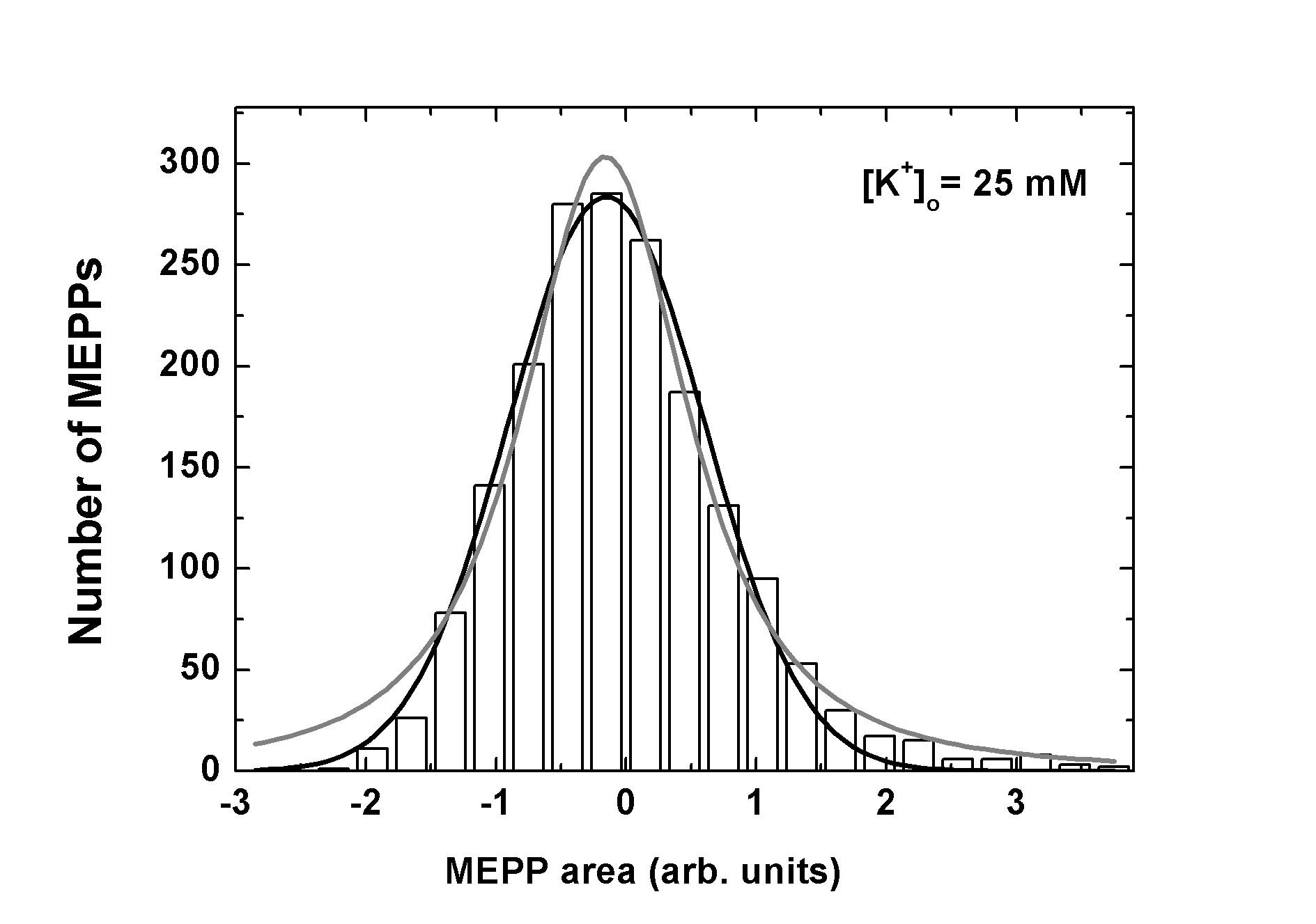}
\caption{\label{figure3} Histograms of MEPPs areas from recordings at $[K^{+}]_{o}$= 5 mM and 25 mV, where gray lines correspond to \textit{q}-Gaussian curves (\textit{q} = 1.45 for $[K^{+}]_{o}$= 5 mM and \textit{q} = 1.7 for $[K^{+}]_{o}$= 25 mM), while black lines represent Gaussian functions.}
\end{figure}

In figure~\ref{figure3} we plot histograms of MEPP areas for $[K^{+}]_{o}$= 5 mM and 25 mM. \textit{q}-gaussian curves were fitted using parameters extracted from ML\textit{q}E estimators, providing a better fit than that of a gaussian distribution. It is important to point out that, before the estimation of the \textit{q}-index from recordings, we verified that all experimental data followed a symmetric distribution, suitable for fitting with gaussians or \textit{q}-gaussians.

\begin{figure}[ht]
\includegraphics[width=8cm]{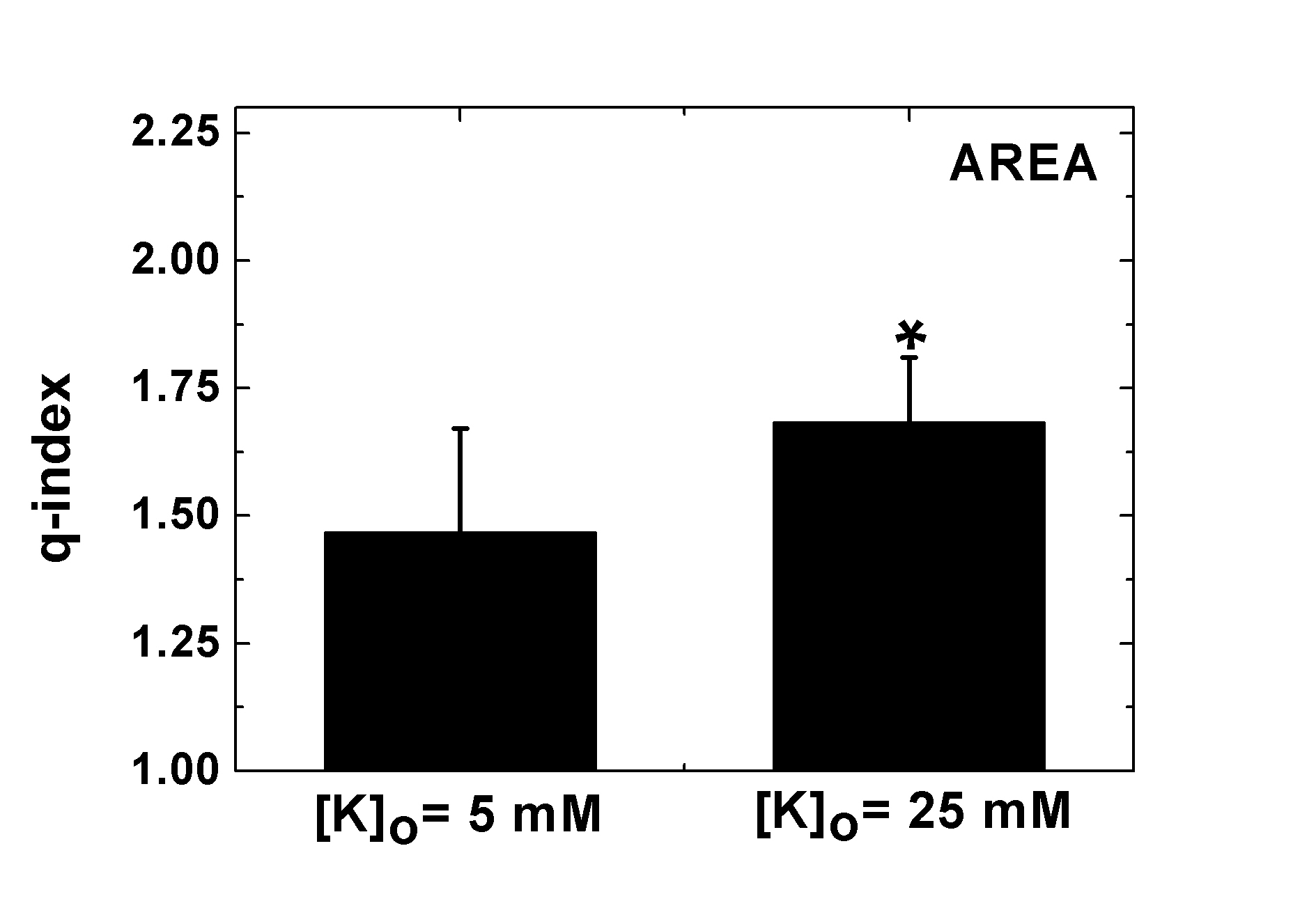}
\includegraphics[width=8cm]{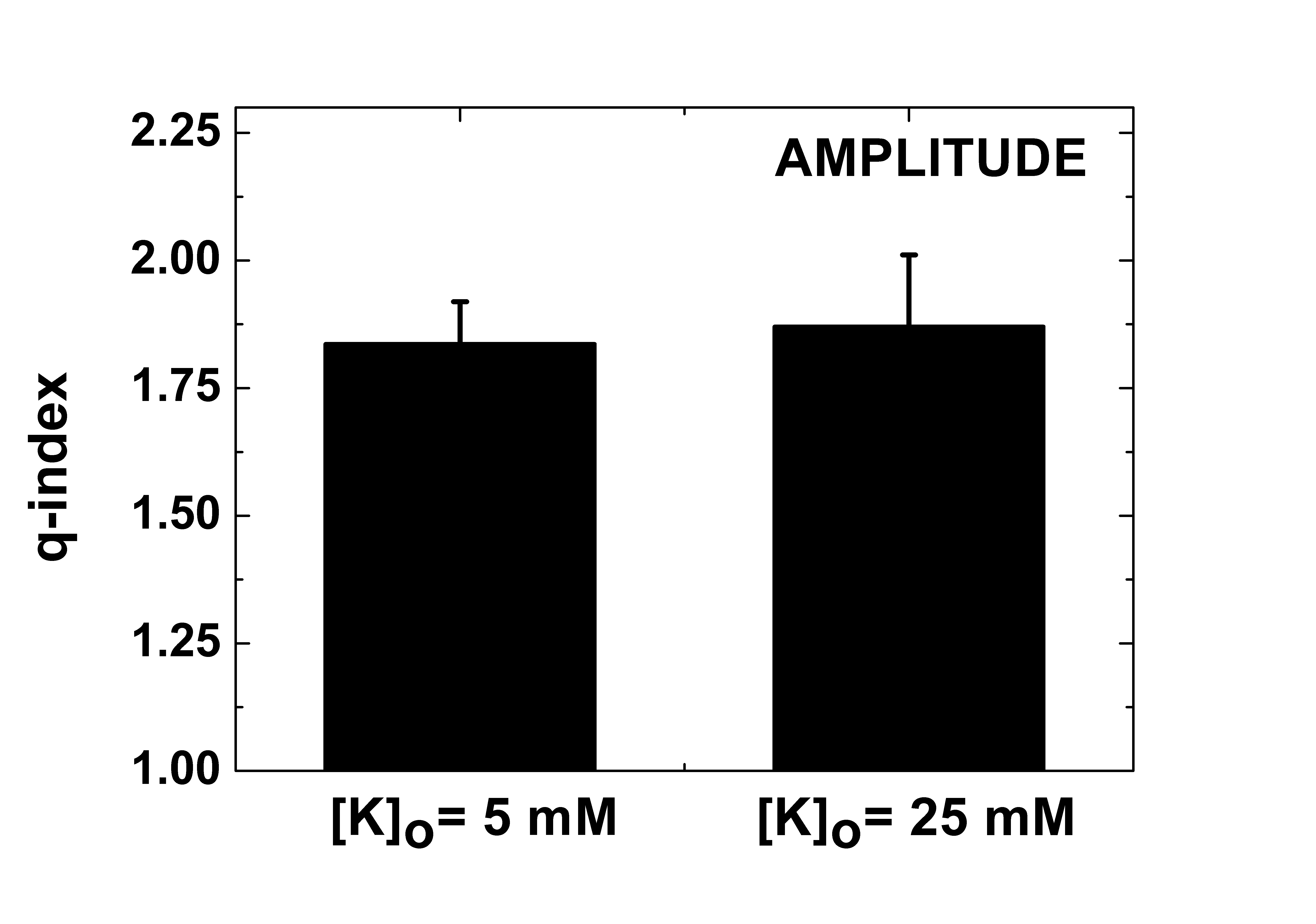}
\caption{\label{figure4} Statistical summary of the results for MEPP areas and amplitudes. For areas, n = 8 for both $[K^{+}]_{o}$. For amplitudes, n = 10 ($[K^{+}]_{o}$= 5 mM) and n = 8 ($[K^{+}]_{o}$= 25 mM). The \textit{q}-index was calculated with the q-estimator(equations(\ref{deriv1} and~\ref{deriv2}). The asterisk highlights a significant difference in the entropic index estimated using MEPP area (p$=$0.024).}
\end{figure}

The amplitudes and areas associated to each event were collected and used to calculate the \textit{q}-index, as can be seen in the statistical summary shown in the figure~\ref{figure4}. The results, expressed as mean $\pm$ standard deviation, were tested for statistical significance using an unpaired t-test and Lillierfors pos-test (p$=$0.024). A significant statistical difference was only observed in MEPP area, as quantified by an elevation in the average \textit{q}-index. Only a slight non-significant increase was observed in MEPP amplitudes.
In table~\ref{table2} we show a summary of \textit{q}-indexes for both concentrations where it is also noteworthy that the confinement effect of \textit{q}-index observed in normal $[K^{+}]_{o}$ increases at high concentration.
\begin{table}[ht]
\caption{\label{table2} Values for \textit{q}-indexes calculated at both potassium concentrations.}
\begin{tabular}{@{}ccccc@{}}
\hline
                   & \multicolumn{2}{c}{Amplitude}  & \multicolumn{2}{c}{Area}     \\ \cmidrule(l){2-5}
                   & 5 mM            & 25 mM         & 5 mM          & 25 mM         \\ \midrule
\textit{q}-index   & $1.83 \pm 0.14$ & $1.87\pm0.08$ & $1.47\pm0.20$ & $1.68\pm0.13$ \\
\textit{q}-minimum & 1.58            & 1.75          & 1.29          & 1.49          \\
\textit{q}-maximum & 2.04            & 2.02          & 1.82          & 1.86
\end{tabular}
\end{table}
\section{Discussion}

In the present work, first we developed calculations to estimate nonextensive indices by application of ML\textit{q}E; then, we applied the theoretical results to electrophysiological data taken from NMJ submitted to an increase of excitability, under elevated $[K^{+}]_{o}$. These two complementary approaches allow us to formulate three general questions: (1) Which biological mechanisms are associated with nonextensivity changes during hyperexcitability? (2) Can these mechanisms be used to characterize the heterogeneity of release probability of neurotransmitters? (3) Why \textit{q}-index values are restricted or confined in this physiological system? To address these issues, we will discuss a possible general scenario by relating physiological mechanisms and nonextensive statistics.

The elevation of $[K^{+}]_{o}$ in NMJ produced a expected increase in MEPP frequency allowing an exame of potential modifications at the pre-synaptic level. Previous studies established a connection between structural or morphological changes and the electrophysiological profile promoted by increase of $[K^{+}]_{o}$. In a study in the frog NMJ using similar approach, Cecarrelli \textit{et al.} documented an increase of MEPP frequency followed by morphological alterations such as dimples and protuberances between active zones \cite{cecarrelli}. The authors also pointed out that $[K^{+}]_{o}$ first triggers vesicular release at the active zones, suggesting that this ion acts in synergy with the exocytosis machinery \cite{cecarrelli2}. From their results, it is worth that the size of active zones and the separation between them can vary with high $[K^{+}]_{o}$. On the other hand, microscopy studies showed that vesicle diameter remains unaffected in this modified medium during \textit{K}-propionate treatment \cite{florey}. Structural studies also revealed the presence of more than a single vesicle in each active zone, as well as interactions between zones even at physiological ionic concentrations. We speculate that due to vesicle volume invariance and nerve ending expansion, such organelles could have more available space in the nerve terminal. This enlarged environment would accommodate  more vesicles than at normal $[K^{+}]_{o}$. This rise in vesicle number results in a nonextensivity increase as it produces strengthening of long-range interactions. Due to its heavy tail property, of the q-Gaussian function can be helpful because it can incorporates classes of vesicles that are neglected by Gaussian fitting (Lupa. 1987). This statistical model could thus be especially useful to shed a new light on the nature of the giant and sub-miniature end plate potentials observed at the NMJ \cite{bennett1995}.

Another possible mechanism is the increase of the tortuosity of the NMJ extracellular medium caused by high $[K^{+}]_{o}$ \cite{nicholson}. A study by Lacks suggests a physical basis for anomalous diffusion as dependent on the tortuosity of the NMJ, whereas Frank proposed a general link between anomalous diffusion and nonextensivity \cite{lacks,frank}. Beyond the anomalous diffusion identified in the extracellular portion of the NMJ, the existence of crowded vesicles and networks of filaments in an inhomogeneous molecular assembly is also potential substrate for anomalous diffusion \cite{norrelykke,santamaria,stauffer,potokar}. Consequently, nonextensivity may also be regulated by the anomalous transport of vesicles in the cytoplasm orchestrated by structural changes in the extracellular space induced by high $[K^{+}]_{o}$.

Accumulation of $[K^{+}]_{o}$ in the CNS is partially involved in the biochemical mechanisms of neurotoxicity related to cerebral ischemia, brain trauma, and inflammation \cite{chang}. For example, this ion can reach concentrations of about 60 mM in the extracellular milleu during spreading depression \cite{Obrenovitch}. The hyperexcitability would modify the statistical profile expressed by the increment in nonextensivity. Therefore, we believe that ML\textit{q}E estimators could emerge as an method to identify nonextensivity in pathological cases related to elevation of $[K^{+}]_{o}$.

The mechanistic steps involved in neuroplasticity and the statistical basis of neurotransmitter secretion are still a matter of debate \cite{neher1}. Despite its limitations, binomial statistics are broadly assumed as the pillar to quantify the neurotransmitter release \cite{neher2}. Modification imposed to the NMJ morphology, emerges as a preponderant synaptic plasticity regulator as suggested by comparative anatomy and electrophysiological investigations \cite{slater,banker}. In this framework, a connection between $[K^{+}]_{o}$, nonextensive statistics, and mechanisms involved in plasticity can be discussed. Release probability and plasticity strength are intimately related \cite{castellucci}. From this, if high $[K^{+}]_{o}$ has the ability to disturb the release of neurotransmitters through electrical and geometrical changes in the terminal, then the probability of release and synaptic gain will be susceptible to structural reorganization themselves.

Potassium channels have been recognized in the pre-synaptic terminal in the NMJ and in CNS synapses. Indeed, a study by Huang and Trussel reported evidences for the importance of this ion in regulating release probability in the rodent brain stem \cite{huang}. The authors used a combination of electrophysiology and immunohistochemistry to reveal the KCNQ5 channel as the controller of synaptic strength in the giant synapse of the Calyx of Held.  Hence, the presence of potassium conductances at the terminal suggest the possibility for an ionic substrate associated to the nonextensive modulation.  It is useful to stress that $[K^{+}]_{o}$ fortify the vesicular release at the nervous terminal \cite{cecarrelli}. This effect can exacerbates the heterogeneous release release given by the lateral inhibition and multiquantal discharges \cite{bennett2,triller}. These interactions in the synaptic terminal can be understood as responsible for long-range correlations, where heterogeneity in the probability of release emerges as a phenomenon explainable by nonextensive statistics. Thus, we speculate that potassium channels close to the active zones underlie, at least partially, the enhancement the enhancement of nonextensivity, as they participate in the interactions mentioned before. Regarding this, it is essential to identify which type of potassium conductance expressed in the NMJ terminal is responsible for the interplay between this ion and neurotransmitter secretion in absence of afferent electrical stimulation.

A statistically significant difference between \textit{q}-indexes was detected only with MEPP area analysis (figure~\ref{figure4}). We attribute the lack of difference between relative amplitudes to the higher susceptibility of this parameter to microelectrode position \cite{tremblay,klootreview}. This effect is pronounced in a situation involving cell swelling, where the NMJ structure/geometry is modified by high $[K^{+}]_{o}$. For this reason, MEPP amplitude, in spite of its customary adoption in electrophysiological studies, can incorporate artifacts by exogenous agents or occasional incorrect experimental methodology. However, a more detailed investigation of amplitude analysis is still necessary to more accurately access its relation with nonextensivity.

Although the entropic index lies in the range 1$<\textit{q}<$3, the results from MEPP areas and amplitudes showed that \textit{q}-indexes belong to a narrower interval (confinement), even at normal $[K^{+}]_{o}$, reinforced under high concentration of this ion (table~\ref{table2}). Such restriction was already observed experimentally in atomic transport of dissipative optical lattices by Douglas \textit{et al.} \cite{douglas}. Along these lines, Bagci and Tirnakli presented a theoretical explanation for the confinement through of a generalized version of Klimontovich S-Theorem \cite{bagci,klimo}. They attributed this behavior to the renormalization of the effective energy. Obviously, neuromuscular dynamics is an example of a complex system interpreted according to classical physics laws. In our particular case, an increased confinement in the \textit{q}-index values would require, beyond high $[K^{+}]_{o}$, temperatures below the physiological level. Many authors reported changes in NMJ electrophysiological properties caused by both hypothermia and hyperthermia exposition \cite{kloot,carlson,nishimura,lilit}. Such investigations found lower MEPP rates under room temperature (24$\pm$1$^\circ{}$C) as compared to physiological temperature (37$\pm$1$^\circ{}$C). Since MEPP shape carries a 'fingerprint' of temperature elevation, we hypothesize that a large variation of the \textit{q}-index would also occur in a warmer environment. At the same time, physiological temperature variation would favor the augment of long-range correlations in the vesicular secretion. According to this view, hyperthermia produces an enormous acceleration of vesicular release, also inducing an enhancement of both variance of \textit{q}-index values and probability for interaction between vesicles. In this scheme, more neighbor vesicles would share Soluble NSF Attachment Protein Receptor (SNARE) proteins placed in the same active zone. As a consequence, this condition would be propitious for a reinforcement of lateral inhibition and multiquantal secretion.

Finally, it is valid to mention a possible role of membrane fluctuations in the nonextensivity increment by using a thermodinamical view. Procopio and Forn\'es employed the fluctuation-dissipation theorem to show that voltage fluctuations regulates the gating behavior of ionic channels \cite{procopio}. In addition, a relation between $[K^{+}]_{o}$ and resting potential fluctuability was recognized by Denker and Poussard, whereas Oosawa argued in favor of coupling between MEPP discharge rate and random membrane potential fluctuations \cite{derksen,poussard,oosawa}. From these studies we suggests that higher temperature and $[K^{+}]_{o}$ intensifies membrane voltage fluctuations, increasing the MEPP rate, forming a propitious enviroment for the exacerbation of a nonextensive pattern. Nevertheless, enquiries need to be considered to clarify these interpretations at physiological temperature and during high $[K^{+}]_{o}$ perfusion.

\section{Conclusion}
In summary, high excitability and morphological changes seem to be prerequisites for the modulation of long-range correlation in nerve endings. To the best of our knowledge, this is the first report to evaluate the influence of $[K^{+}]_{o}$ in the neurotransmission statistics. Many reports have stressed the role of potassium accumulation as a primary agent in brain diseases. Nonextensivity can be inherent to diseases specifically associated to hyperexcitability promoted by elevated concentrations of this ion. Tsallis formalism poses as an alternative tool to describe neurotrasmitter release statistics in conditions such as denervation and osmotic stress at the NMJ or spreading depression and epileptiform activity in the CNS.  It is possible to extend the present strategy to CNS synapses, such as those involved in excitatory and inhibitory transmission in the brain. It would also be important to perform \textit{in vitro} investigations during ischemia and epileptiform activity. Moreover, according to our point of view, lateral inhibition and multiquantal release are dominant phenomena at normal level of excitability, but they can be magnified during due to high $[K^{+}]_{o}$. We expect that a detailed theoretical model associating long-range correlations and heterogeneity of release probability inspired by nonextensive statistics will be developed. Also, post-synaptic contributions to nonextensivity remain to be studied in detail. Finally, we understand that additional investigations on the statistical properties of the estimator here introduced are necessary.

\begin{acknowledgements}
This work was supported in part by CNPq. The authors would like to thank to Constantino Tsallis for valuable comments and Christopher Kushmerick for his support.
\end{acknowledgements}

\bibliography{text2refer}

\end{document}